\documentclass[prd,letterpaper,tightenlines,
showpacs,preprintnumbers,nofootinbib,floatfix,psfig]{revtex4}

\usepackage{amsmath,amssymb,amsfonts,graphicx,pifont,dcolumn}
\usepackage{epstopdf}
\usepackage{color}
\usepackage{cancel}
\usepackage{ulem}
\RequirePackage{slashed}

\begin{document}

\title{Ion scattering on monopoles}

\author{Vicente Vento} 

\affiliation{Departamento de F\'{\i}sica Te\'orica-IFIC, Universidad de Valencia- CSIC, 46100 Burjassot (Valencia), Spain.}

\date{\today}

\begin{abstract}
Magnetic monopoles have been a subject of interest since Dirac established the relation between the existence of monopoles and charge quantization. The Dirac quantization condition bestows the monopole with a huge magnetic charge. We study the scattering of charged ions by monopoles and use backscattering techniques to devise a method to detect monopoles bound in matter.
\end{abstract}

\pacs{03.65.Pm, 13.40.-f, 14.80.Hv}

\maketitle

\section{Introduction}
The theoretical justification for the existence of classical magnetic poles, hereafter called
monopoles, is that they add symmetry to Maxwell's equations and explain charge quantization. Dirac showed that the mere existence of a monopole in the universe could offer an explanation of the discrete nature of the electric charge.
His analysis leads to the Dirac Quantization Condition (DQC) \cite{Dirac:1931kp,Dirac:1948um}

\begin{equation}
e g = N/2, \; N = 1,2,...,
\label{dquatization}
\end{equation}
where $e$ is the electron charge, $g$ the monopole magnetic charge and we use natural units
$\hbar= c=1= 4 \pi \varepsilon_0$. Monopoles have been a subject of experimental interest since Dirac first proposed them
in 1931.

In Dirac's formulation, monopoles are assumed to exist as point-like particles and quantum mechanical consistency conditions lead to establish the value of their magnetic charge.  
Due to the large magnetic charge as a consequence of Eq.(\ref{dquatization}) monopoles can bind in matter \cite{Milton:2006cp}. In the case of the Dirac monopole theory the other monopole parameter is its mass. 
Searches for direct monopole production have been performed in most accelerators.
The lack of monopole detection has been transformed into  monopole mass lower bounds 
\cite{Abulencia:2005hb,Fairbairn:2006gg,Abbiendi:2007ab,Aad:2012qi}. 
Experiments at LHC have probed higher masses \cite{Aad:2015kta,Lenz:2016zpj,Acharya:2014nyr,MoEDAL:2016jlb,Acharya:2016ukt,Acharya:2017cio}.  In here we shall use that the monopole mass is larger than 400 GeV.

Since monopoles are stable after formation they may bind into conventional materials like beam pipes, other detector elements, the beam dump material  or magnetic monopole trapping volumes as in the MoEDAL experiment \cite{Acharya:2014nyr,MoEDAL:2016jlb,Acharya:2016ukt,Acharya:2017cio}. The aim of this paper is to study the collision of charged ions with bound monopoles to obtain signatures for monopole detection. However, since the monopole will be surrounded in practice by a order Avogadro's number of conventional nuclei, to isolate the monopole signal from  conventional ion-nucleus Rutherford scattering, we propose the use of back scattering techniques.

\section{Scattering of a charged spin $1/2$ particle by a spinless magnetic monopole.}

\begin{figure}[htb]
\begin{center}
\includegraphics[scale= 0.9]{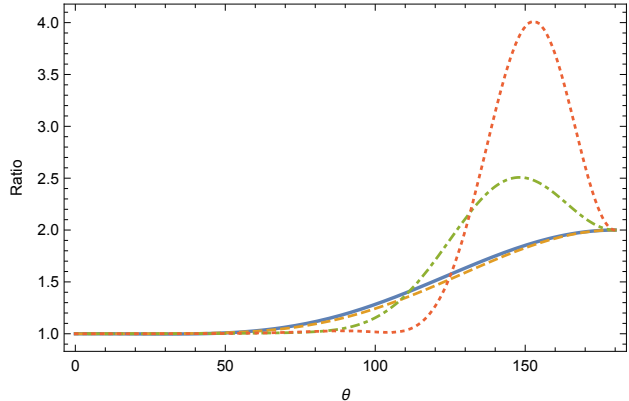} 
\end{center}
\caption{We plot the function  $R(q,\theta)$ defined in Eq. (\ref{eqratio})  for $q=1/2$ (solid), q=2 (dashed), q=4 (dotdashed) and $q=6$ (dotted).}
\label{figratio}
\end{figure}

Let us assume the following scenario a beam of charged particles is scattered on a monopole bound in matter. For the time being we omit the scattering of the particles on the background and study the problem of the scattering of a relativistic particle by the central potential of a monopole field. In a beautiful paper Kazama, Yang and Goldhaber described the scattering of a relativistic spin 1/2 particle with charge $Ze$ by a fixed spinless magnetic monopole \cite{Kazama:1976fm}. The formulation developed in terms of fiber bundles is absent of the string singularity obtains the result by the use of monopole harmonics \cite{Wu:1976ge}.  The result for the differential cross section for an unpolarized beam is given by

\begin{equation}
\frac{d\sigma}{d\Omega} = \frac{1}{2 k^2} \left(|T_{|q|}|^2 + 2 q^2 (\sin{(\theta/2)}^{4|q|-2}\right),
\end{equation}
where $q = Zeg = Z/2$, $\theta$ is the scattering angle and the function $T_{|q|}$ is a complex expression defined by Eq.(80) of ref. \cite{Kazama:1976fm}. Its behavior is similar for low $q$ to the conventional Rutherford formula as can be seen in Fig. \ref{figratio} where we plot

\begin{equation}
Ratio (q,\theta)=\left( \frac{Z' e}{g v}\right)^2 \left(\frac{d\sigma}{d\Omega}{\Big / }{\frac{d\sigma}{d\Omega}_R}\right)
\label{eqratio}
\end{equation}
for a beam of particles of velocity $v$, charge $Ze$ scattered by a target particle of charge $Z'e$ for f $q=0.5, 2, 4, 6 $. This ratio is independent of beam momentum. The ratio varies dramatically with $Z$ specially in the backward direction where the cross section is smallest.

\begin{figure}[htb]
\begin{center}
\includegraphics[scale= 0.9]{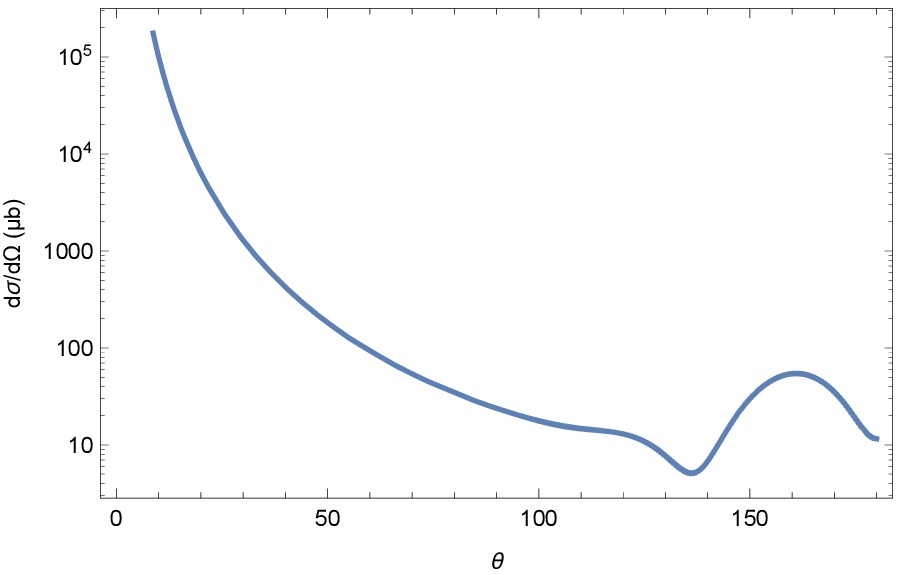} 
\end{center}
\vskip -5.35cm\hskip 3.7cm \includegraphics[scale= 0.45]{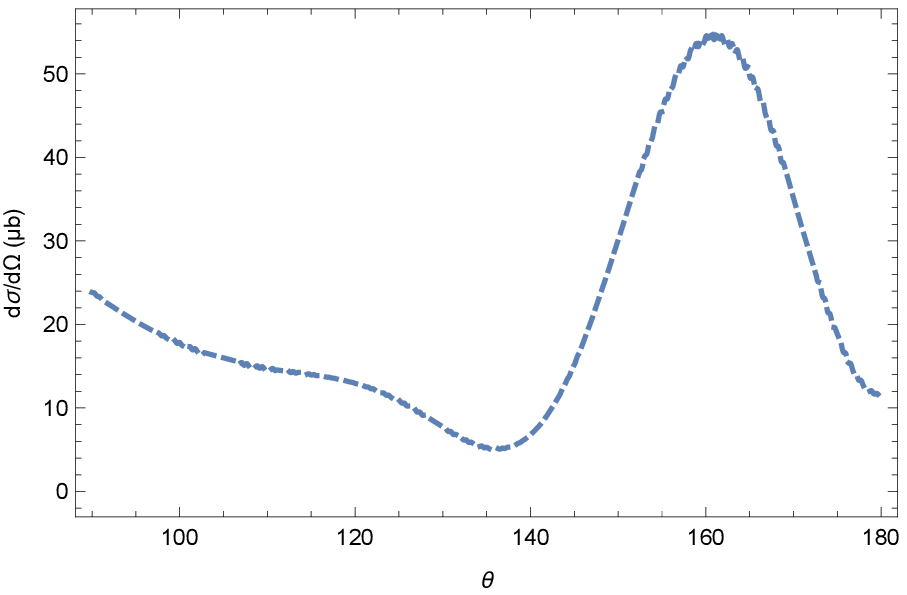} 
\vskip 3cm
\caption{Elastic scattering ion monopole cross section ($\mu$ b)  a for a beam energy  of 1 GeV/nucleon spin $1/2$ ions (Z= 20, A=41)  by a fixed spinless magnetic monopole. The inset  shows in dashed the details of the backward region of experimental interest.} 
\label{sigma}
\end{figure}

In Fig. \ref{sigma} we plot the differential cross section for a beam of 1 GeV/nucleon ions of $Z=20, A=41$. We note that the presence of a monopole close to the beam makes the probability of scaterred ions coming out transverse and backward to the beam measurable.

\begin{figure}[htb]
\begin{center}
\includegraphics[scale= 0.9]{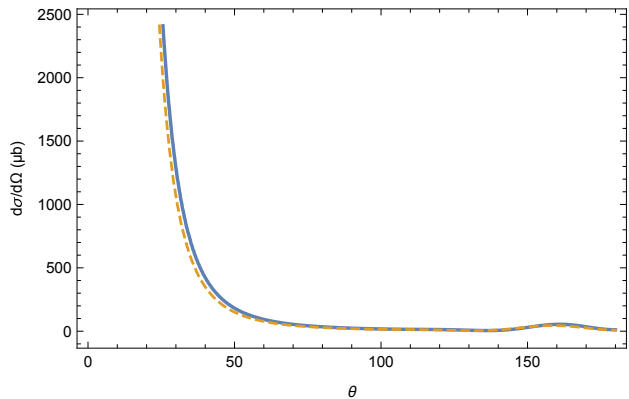} 
\end{center}
\caption{Elastic scattering cross section ($\mu$ b) for a  $1$ GeV/nucleon beam of spin $1/2$ ions (Z= 20,A=41)  by a monopole of infinite mass (solid) and for  monopole of mass $M= 10 m_A$.} 
\label{labsigma}
\end{figure}

Moving from this approximation of infinitely massive monopoles to monopoles with a finite mass does not change the result much, specially since the expected masses of the monopole is supposed to be greater the 400 GeV \cite{Abulencia:2005hb,Fairbairn:2006gg,Abbiendi:2007ab,Aad:2012qi,Acharya:2016ukt}. We plot in Fig. \ref{labsigma} for a heavy ion of $Z=20, A=41$ the comparison between the infinite and finite mass case, the latter governed by the equation \cite{Parzen:1950pa},

\begin{equation}
\frac{d\sigma}{d\omega}(\theta) = \left(2 \mu \cos{\theta} + \frac{1+ \mu^2 \cos {2\theta} }{\sqrt{1-\mu^2 \sin^2{\theta}}}\right)\frac{d\sigma}{d\Omega}(\theta +\arcsin(\mu \sin \theta))
\end{equation}
where the lower case variables correspond to the lab frame and $\mu = m_A/M$, $m_A$ being the charged fermion mass and $M$ the monopole mass. 

We might conclude this discussion by stating that a monopole sends beam particles of known momentum (elastic scattering) into the transverse and backward directions which are  easy to detect. We also notice that the structure in the backward direction as shown in Fig. \ref{figratio} can be very different from the traditional Rutherford scattering. Those two features are clear signals of the existence of bound monopoles.

\section{Backscattering spectrography}

The ion-monopole cross section is large, however since the probes are macroscopic, the monopoles will be surrounded  by a huge number, $\sim 10^{24}$, of conventional scatterers and therefore the background might hide the monopole signal. 

\begin{figure}[htb]
\begin{center}
\includegraphics[scale= 0.9]{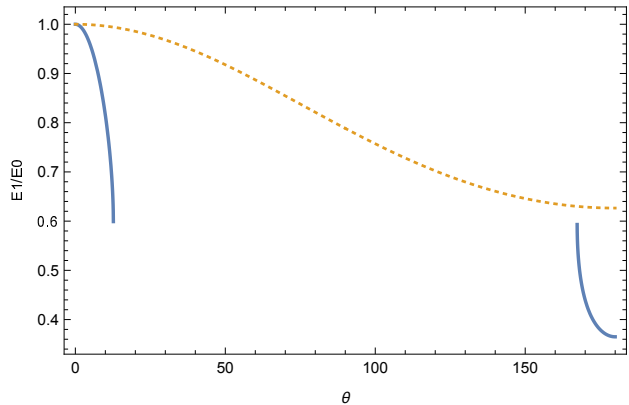} 
\end{center}
\caption{Kinematical factor for $^{41}_{20}$Ca on $^9_4$Be at beam energy of 1GeV/nucleon(solid), and for $^{41}_{20}$Ca on a monopole whose mass is taken to be $10 m_{41}$.} 
\label{backscattering}
\end{figure}

To avoid this problem we recall the Rutherford backscattering technique of condensed matter physics\cite{Feldman} . We consider the elastic scattering of a particle in the beam with mass $m_b$ and a stationary particle of mass $m_t$ located in the sample. Let us consider the kinematics of a non-relativistic collision with $m_b > > k$ we recall that the energy of the scattered projectile $E_f$ is reduced from the initial energy $E_i$ by the so called kinematical factor $E_f=\kappa E_i$, where

\begin{equation}
\kappa=\frac{m_b \cos{\theta_L} \pm \sqrt{ m_t^2 -m_b^2 (\sin{\theta_L})^2}}{m_b +m_t},
\label{kappa}
\end{equation}
where $\theta_L$ is the scattering angle of the projectile in the laboratory frame. The plus sign is taken when the mass of the projectile is less than that of the target, otherwise the minus sign is taken. Eq.(\ref{kappa}) is a consequence of energy momentum conservation. This equation shows that if $m_b > m_t$ for certain angles the square root admits no real solution and therefore there is an angular sector were scattering is strictly forbidden by energy momentum conservation. We plot the $\kappa$ factor in Fig.\ref{backscattering}.

The generalization to relativistic kinematics of Eq.(\ref{kappa}) is straightforward but less transparent for visualizing the kinematical restriction \cite{oravavento}. Thus, scanning the probe with a beam of particles whose atomic number is greater than any atomic number in the probe there will be no background in a large angular region and therefore we will obtain a clean signal if monopoles are bound in the probe. 

 \section{Conclusion}
 
The discovery of monopoles would be a major breakthrough in our understanding of charge quantization and would imply revisiting Quantum Electrodynamics in a strong coupling regime. Monopoles are stable particles and they bind to conventional matter, specifically  to the nuclei of atomic systems \cite{Milton:2006cp}. Several techniques have been used to detect monopoles, in particular the most common is to measure the magnetic charge of matter probes with magnetometers \cite{MoEDAL:2016jlb,Acharya:2016ukt,Acharya:2017cio}. We have shown here that ion beams can be used to detect monopoles, by discussing an effective way of eliminating the conventional background using the same technique that condensed matter physicists use to detect impurities  \cite{Feldman} . 

Given the lack of a fundamental theory with well known parameters for monopole production it is very unlikely that our technique might substitute the much cheaper use of magnetometers. However, one can use monopole detection via ion scattering at end of the line experiments without or with little additional equipment. For example, a monopole bound to the beam pipe of an accelerator,  can be detected by the set up used for Central Exclusive Production (CEP) \cite{oravavento,McNulty:2017ejl} by carrying out a van der Meer sweeping with the ion beam, or if bound in the material of the beam dump it can be detected by locating a few detectors for beam particles backward of the beam impact region.

We have shown that the large coupling constant and the mass of the monopoles single out this particle with respect to the abundant conventional particles and therefore by an intelligent use of ion scattering end of the line experiments monopoles would be detectable if bound to matter.

\section*{Acknowledgments}
I thank Fernando Mart\'{\i}nez, Philippe Mermod and Risto Orava for discussions. 
This work was supported in part by Mineco and UE Feder under contract FPA2016-77177-C2-1-P and SEV-2014-0398.

\end{document}